\documentclass[12pt]{article}
\usepackage{graphicx}
\newcommand{\beq}{\begin{equation}}
\newcommand{\eeq}{\end{equation}}
\newcommand{\bea}{\begin{eqnarray}}
\newcommand{\eea}{\end{eqnarray}}

\def\lap{\mathrel{\mathpalette\fun <}}

\def\fun#1#2{\lower3.6pt\vbox{\baselineskip0pt\lineskip.9pt
  \ialign{$\mathsurround=0pt#1\hfil##\hfil$\crcr#2\crcr\sim\crcr}}}
\begin{document}
\begin{titlepage}
\begin{flushleft}
       \hfill                      {\tt hep-th/0204066}\\
       \hfill                       FIT HE - 02-01 \\
\end{flushleft}
\vspace*{3mm}
\begin{center}
{\bf\LARGE Localization of Gravity on Brane Embedded
in $AdS_{5}$ and $dS_{5}$  \\ }
\vspace*{5mm}

\bigskip

{\large Iver Brevik \footnote{\tt iver.h.brevik@mtf.ntnu.no}\\}
\vspace{2mm}
{
Division of Applied Mechanics, Norwegian University of Science and Technology,
N-7491 Trondheim, Norway\\}
\vspace*{5mm}

{\large Kazuo Ghoroku\footnote{\tt gouroku@dontaku.fit.ac.jp}\\ }
\vspace*{2mm}
{
${}^2$Fukuoka Institute of Technology, Wajiro, Higashi-ku}\\
{
Fukuoka 811-0295, Japan\\}
\vspace*{5mm}

{\large Sergei D. Odintsov\footnote{\tt odintsov@mail.tomsknet.ru}\\ }
\vspace*{2mm}
{
${}^3$ Tomsk State Pedagogical University, 634041 Tomsk, Russia}\\
\vspace*{5mm}

{\large Masanobu Yahiro \footnote{\tt yahiro@sci.u-ryukyu.ac.jp} \\}
\vspace{2mm}
{
${}^4$Department of Physics and Earth Sciences, University of the Ryukyus,
Nishihara-chou, Okinawa 903-0213, Japan \\}

\vspace*{10mm}

\end{center}

\begin{abstract}
We address the localization of gravity on the Friedmann-Robertson-Walker
type brane embedded in either $AdS_{5}$ or $dS_{5}$ bulk space,
and derive two definite limits between which the value of the
bulk cosmological constant has to lie 
in order to localize the graviton on the brane. 
The lower limit implies that the brane should be 
either $dS_{4}$ or $4d$ Minkowski in the $AdS_{5}$ bulk.
The positive upper limit indicates that the gravity can be trapped 
also on curved brane in the $dS_{5}$ bulk space. 
Some implications to 
recent cosmological scenarios are also discussed.

\end{abstract}
\end{titlepage}

\section{Introduction}
It is quite expectable that four dimensional world is formed in
the process of compactification
from the ten-dimensional superstring theory. The D-brane
approach is getting very useful for the study of such theory.
In particular, there is some interest
 in the geometry obtained from the D3-brane of type IIB theory.
Near the horizon of the stacked D3-branes, 
the configuration  $AdS_5\times S^5$ is realized and 
the string theory on this background  describes
the four-dimensional SUSY Yang-Mills theory which lives on the
corresponding boundary
\cite{M1,GKP1,W1,Poly1}. 

On the other hand, a thin three-brane (Randall-Sundrum brane)
embedded in $AdS_5$ space has been proposed ~\cite{RS1,RS2} as 
a model of our world.
The position of the brane is possible at any 
value of the transverse coordinate, which is considered as the
energy scale of the dual conformal field theory (CFT) living on the boundary. 
Another interesting point is that
this idea gives an alternative to the standard Kaluza-Klein (KK)
compactification via the localization of the zero mode of the graviton 
\cite{RS2}.
Brane approach opened also a new way to the construction of
the hierarchy between four-dimensional Planck mass and
the electro-weak scale, and also for realization of the small
observable cosmological constant
 with lesser fine-tuning \cite{ArHa,KaSc}.

The gravity theory under consideration is  five-dimensional. However, its
zero mode is trapped on the brane, and as a result the 
usual 4-dim Newton law  is realized on the brane.
Non-trapped, massive KK modes correspond to a 
 correction to Newton's law, and they are also understood
from the idea of the
AdS/CFT correspondence. 
Localization of the various fields, especially of gravity, is thus essential 
for the brane-world to be realistic. 
The study of localization, however, was limited to the case 
of $AdS_{5}$ bulk space and not made for $dS_{5}$ bulk space.

The purpose of this paper is to
study the localization of the graviton on our brane 
when it is taken to be time dependent and 
embedded in  $AdS_5$ or in $dS_5$.
In Section 2, we give various brane solutions obtained on the basis of 
a simple ansatz imposed on the bulk metric. For these solutions, 
we make some  brief comments from a cosmological viewpoint.
In Section 3, the localization of the graviton on those
brane solutions are examined, and a restriction for the parameters
of the theory is given  in order to realize the localization (cf. Eq.~(61)
below). 
In Section 4, the relation of our solutions to cosmology, 
especially to the recently observed mini-inflation, 
is discussed. 
Also, the sub-class of viscous cosmology
is briefly dealt with. 
Concluding remarks are given in the final section.

\section{Cosmological solutions of brane-universe}

We start from the five-dimensional gravitational action. It is given in the
Einstein frame as\footnote{
Here we take the following definition, $R_{\nu\lambda\sigma}^{\mu}
=\partial_{\lambda}\Gamma_{\nu\sigma}^{\mu}-\cdots$, 
$R_{\nu\sigma}=R_{\nu\mu\sigma}^{\mu}$ and $\eta_{AB}=$diag$(-1,1,1,1,1)$. 
Five dimensional suffices are denoted by capital Latin and the four
dimensional
one by the Greek ones.
}
\beq
    S_5 = {1\over 2\kappa^2}\Bigg\{
      \int d^5X\sqrt{-G} (R -  2\Lambda + \cdots)
          +2\int d^4x\sqrt{-g}K\Bigg\}, \label{action}
\eeq
where the dots denote the contribution from matter,
$K$  being the extrinsic curvature on the boundary. This term is formal here
and it plays no role until one considers the AdS/CFT correspondence.
The fields represented by the dots
 are not needed to construct the background 
of the brane. 
The other ingredient is the brane action,
\beq
    S_{\rm b} = -{\tau}\int d^4x\sqrt{-g}, \label{baction}
\eeq
which is added to $S_5$, and the Einstein equation is written as
\beq
 R_{MN}-{1\over 2}g_{MN}R=\kappa^2T_{MN} \label{equation}
\eeq
where $\kappa^2T_{MN}=-(\Lambda+{1\over b}\delta(y)\kappa^2\tau
\delta_{\mu}^M\delta_{\nu}^N) g_{MN}$ and $b=\sqrt{-g}/\sqrt{-G}$.
Here we solve the Einstein equation (\ref{equation}) with the following
metric,
\beq
 ds^2= -n^2(t,y)dt^2+a^2(t,y)\gamma_{ij}(x^i)dx^{i}dx^{j}
           +dy^2  \, \label{metrica},
\eeq
where the coordinates parallel to the brane are denoted by $x^{\mu}=(t,x^i)$,
 $y$ being the coordinate transverse to the brane. The position of the brane
is taken at $y=0$.

Although the form (\ref{metrica}) is simple, it
could describe various geometries of both the bulk and the 
brane.  The solutions are controlled by  two
parameters, $\Lambda$ and $\tau$. The 
configuration of the bulk is determined by $\Lambda$, i.e., anti-de Sitter
(AdS) space for $\Lambda<0$ and de Sitter (dS) space
for $\Lambda>0$. The geometry of the four-dimensional 
brane is controlled by
the effective four-dimensional cosmological constant, $\lambda$, which follows
from Eq.~ (\ref{Einsteina0}) below and is given explicitly by
\beq
   \lambda = \kappa^4\tau^2/36+\Lambda/6 . \label{4cos}
\eeq 
Thus $\lambda$ is determined by both $\Lambda$ and
the intrinsic four-dimensional cosmological constant, $\tau$.
Here we notice that there are two values of
$\tau=\pm|\tau|$ for the same $\lambda$, but positive $\tau$ should be
chosen for the localization of the gravity since one needs attractive
$\delta$-function force (see Eqs.~(\ref{pot2}) and (\ref{pot3})). 
We restrict our interest here to the case of a Friedmann-Robertson-Walker type
(FRW) universe. In this case, the three-dimensional metric $\gamma_{ij}$
is described in Cartesian coordinates as
\beq
  \gamma_{ij}=(1+k\delta_{mn}x^mx^n/4)^{-2}\delta_{ij},  \label{3metric}  
\eeq
where the parameter values $k=0, 1, -1$ correspond to a
 flat, closed, or open universe respectively.
\vspace{.5cm}

We consider the solution of the Einstein equation (\ref{equation}) 
with the ansatz (\ref{metrica}).
This leads to the following equations: 
\beq
 3\{ ({\dot{a}\over a})^2-n^2({{a''}\over a}+({{a'}\over a})^2)
   +k{n^2\over a^2} \}=\kappa^2T_{tt},
\eeq
\beq
 a^2\gamma_{ij}
   \{ {{a'}\over a}({{a'}\over a}+2{{n'}\over n})+2{{a''}\over a}
                   +{{n''}\over n}\}
   +{a^2\over n^2}\gamma_{ij}\{ 
     {\dot{a}\over a}(-{\dot{a}\over a}+2{\dot{n}\over n})-2{\ddot{a}\over a}
                  \}-k\gamma_{ij}=\kappa^2T_{ij},
\eeq
\beq
 3({{n'}\over n}{\dot{a}\over a}-{\dot{a'}\over a})=\kappa^2T_{ty},
\eeq
\beq
 3\{ {{a'}\over a}({{a'}\over a}+{{n'}\over n})-{1\over n^2}
({\dot{a}\over a}({\dot{a}\over a}-{\dot{n}\over n})+{\ddot{a}\over a})
   -{k\over a^2} \}=\kappa^2T_{yy}.
\eeq
Integrating once the 
$(t,t)$ and $(y,y)$ components of the 
Einstein equation with respect to $y$ \cite{Lan}, one gets
\beq
 ({\dot{a}\over na})^2={\Lambda\over 6}+({a'\over a})^2-{k\over a^2}
           +{C\over a^4},  \label{Einstein}
\eeq
where $C$ is a constant of integration. 
Let us consider this constant more closely. The solution of
Eq.~(\ref{Einstein})
 can be given as \cite{BDEL}, 
$$
 a(t,y)=\left\{{1\over 2}(1+{\kappa^4\tau^2 \over 6\Lambda})a_0^2
       +{3C\over \Lambda a_0^2}+
       [{1\over 2}(1-{\kappa^4\tau^2 \over 6\Lambda})a_0^2
       -{3C\over \Lambda a_0^2}]\cosh(2\mu y)\right. \nonumber
$$
\beq
  ~~~~~~~~~~~~~~~~~~~\left.-{\kappa^2\tau \over \sqrt{-6\Lambda}}
                 a_0^2\sinh(2\mu |y|)\right\}^{1/2} , \label{fullsol1}
\eeq
for negative $\Lambda$ where $\mu=\sqrt{-\Lambda/6}$.
For positive $\Lambda$, the solution is given as
$$
 a(t,y)=\left\{{1\over 2}(1+{\kappa^4\tau^2 \over 6\Lambda})a_0^2
       +{3C\over \Lambda a_0^2}+
       [{1\over 2}(1-{\kappa^4\tau^2 \over 6\Lambda})a_0^2
       -{3C\over \Lambda a_0^2}]\cos(2\mu_d y)\right. \nonumber
$$
\beq
  ~~~~~~~~~~~~~~~~~~~\left.-{\kappa^2\tau \over \sqrt{6\Lambda}}
                 a_0^2\sin(2\mu_d|y|)\right\}^{1/2} , \label{fullsol}
\eeq
where $\mu_d=\sqrt{\Lambda/6}$.
In both cases, $a_0(t)= a(t,y=0)$ and $n(t,y)=\dot{a}(t,y)/\dot{a}_0(t)$.
As for $a_0(t)$, its governing equation
can be obtained by considering Eq.~(\ref{Einstein}) at $y=0$
with the boundary condition at $y=0$,
\beq
  {a'(t,0+)-a'(t,0-)\over a_0(t)}=-{\kappa^2\tau\over 3}, \label{bound22}
\eeq
the latter following from integrating Eq.~(7) across the surface $y=0$. We get
\beq
 ({\dot{a}_0\over a_0})^2={\lambda}-{k\over a_0^2}
           +{C\over a_0^4}.  \label{Einsteina0}
\eeq
(cf. Eq.~(5)). 
The solution of this equation is obtained with a new constant of integration, 
$c_0$:
\beq
 a_0(t)={\lambda^{-3/4}\over 2f(t)}
    (\sqrt{\lambda}(f^4(t)-4C)+2k{\lambda}^{1/4}f^2(t)+k^2)^{1/2} ,
\eeq
\beq
         f(t)=e^{\sqrt{\lambda}(t-c_0)} .
\eeq
This solution shows inflationary behavior for positive $\lambda$. It then
describes 
the inflation at the early universe and the mini-inflation at the present 
universe. Further remarks on this point are given in Section 4. 

\vspace{.5cm}

In terms of these general solutions, it is possible to discuss more
closely the meaning of the parameter $C$. For example,
 $C$ can be related to the CFT radiation field energy in a cosmological 
context, from the viewpoint of the AdS/CFT correspondence \cite{Gub,Pad}.
In this sense, the above solution with non-zero $C$ may be important
to see the cosmological dS/CFT correspondence(for the introduction,
see\cite{Str}).

In this general case, it will however be difficult to proceed with the
analysis
of the localization problem of the graviton. We intend to discuss this
general case elsewhere.  Instead, we impose here the following ansatz on
the metric 
(\ref{metrica}):
\beq
 a(t,y)=a_0(t)A(y), \quad n(t,y)=A(y) \label{metrica3}.
\eeq
This turns out to be convenient for our purpose of examining the
localization problem.
This form
is suitable to see the effects of the bulk geometry $A(y)$ on the cosmological
evolution expressed by the scale factor $a_0(t)$.
It is easy to see that the solution of this type can be obtained 
by taking $C=0$ in Eq.~(\ref{fullsol}). Then the solutions obtained in
the form of (\ref{metrica3}) are restricted to the one where the radiation
energy content
of CFT is neglected. We will see, in Section 4,  that this simplification
is justified
 in a realistic cosmological scenario.
Hereafter, we discuss the solutions of the form (\ref{metrica}) with the
ansatz 
(\ref{metrica3}) for FRW brane-universes, with  $k=0, \pm 1$. 

When  considering Eq.~(\ref{metrica3}), we obtain from Eq.~
(\ref{Einstein}) the
following equations when $C=0$:
\beq
  ({\dot{a_0}\over a_0})^2+{k\over a_0^2}=A'^2+{\Lambda\over 6}A^2
          =D,  \label{Einstein2}
\eeq
where $D$ is a constant being independent of $t$ and $y$. 
In view of the boundary condition at the brane position,
\beq
  {A'(0+)-A'(0-)}=-{\kappa^2\tau\over 3}A(0), \label{bound2}
\eeq
one gets
\beq
   D=\lambda
\eeq 
if  $A(0)=1$. The normalization condition
$A(0)=1$ does not affect the generality of our discussion.
In the following, we discuss solutions for the cases where $\lambda$ takes
zero, positive, 
or negative values.

\subsection{Solutions for $\lambda=0$}

When $\lambda=0$,  solutions are available only for $k=0, -1$,
because of Eq.~(\ref{Einstein2}).
The solution for the flat three space ($k=0$) 
 is known as the Randall-Sundrum (RS) brane \cite{RS1,RS2}, 
and is 
\beq
 ds^2= e^{-2|y|/L}\eta_{\mu\nu}dx^{\mu}dx^{\nu}
           +dy^2  \, \label{metrica2}
\eeq
where $\tau=6/(L\kappa^2)$,
$L=\sqrt{-6/\Lambda}$ being the AdS radius.
Here $\lambda=0$ is achieved by the fine-tuning of $\Lambda$ and $\tau$
as given above. This is assured from Eq.~(\ref{4cos}).
The solution (\ref{metrica2}) represents a static and flat brane
situated at $y=0$, and the configuration is taken
to be $Z_2$ symmetric for $y\to -y$. 

Another solution is obtained for $k=-1$, open three space, as
\beq
 ds^2= e^{-2|y|/L}(-dt^2+a_0^2(t)\gamma_{ij}dx^{i}dx^{j})
           +dy^2  \, \label{metrica22}
\eeq
where $\gamma_{ij}$ is taken from Eq.~(\ref{3metric}) with $k=-1$ and
$a_0(t)=\pm t+c_0$ with a constant $c_0$. This solution leads to
the curvature dominated
universe whose  open three space size expands or shrinks linearly
with time. But this universe would not correspond to our present universe,
since recent analyses of the cosmic microwave background 
indicates that our universe is almost flat.

In the evolution of our universe, the cosmological constant $\lambda$ 
may be not zero but positive. It is considered to be large 
in the early inflation epoch and 
tiny in the present epoch. 
Recent observations of Type Ia supernovae and 
the cosmic microwave background 
indicate that our universe is dominated by a positive $\lambda$
\cite{garnavich,perlmutter,wang}. Next, we consider this case.

\subsection{Solutions for $\lambda >0$}

When $\lambda >0$, we obtain time-dependent solutions for
each value of $k=0, 1, -1$. 
When $k=0$, the typical, inflationary brane is obtained, and it
has the following form \cite{Kal,Nih}
\beq
 a_0(t)=e^{H_0t}, \quad A(y)= {\sqrt{\lambda}\over\mu}
       \sinh[\mu(y_H-|y|)] \, \label{metrica4}
\eeq
where $\mu=\sqrt{-\Lambda/6}$, and the Hubble constant is represented as
$H_0=\sqrt{\lambda}$. 
This solution is
obtained for $\Lambda <0$,  $y_H$ representing the position of the horizon,
and
the five dimensional space-time is expressed as
\beq
  ds^2= A(y)^2(-dt^2+e^{H_0t}\delta_{ij}dx^idx^j)+dy^2. \label{metrica5}
\eeq
This solution represents a brane at $y=0$. The configuration is taken
to be $Z_2$ symmetric as in Eq.~(\ref{metrica2}). 
It should be noticed that 
the solution approaches the RS solution (\ref{metrica2}) 
in the limit $H_0\to 0$. In this sense, the solution can be understood as an
extension of the RS brane to a time-dependent brane due to the 
non-zero $\lambda$, which in turn can be considered 
as a result of the failure of the fine-tuning to obtain zero $\lambda$.

For $k=\pm 1$, the solutions are given 
by the same $A(y)$ but with a different $a_0(t)$;
\beq
 a_0(t)={1\over H_0}\cosh(H_0t+\alpha_1)
         \, ,\label{desitp1}
\eeq
for $k=1$ and 
\beq
  a_0(t)={1\over H_0}\sinh(H_0t+\alpha_2)
         \, ,\label{desitm1}
\eeq
for $k=-1$, where $\alpha_i$ are constants. These solutions also
represent  inflation with curved three space.
For any value of $k=0, \pm1$,
we obtain the same solution for $A(y)$ as given above. Thus $a_0(t)$ has
nothing to do with the problem of
 localization since it depends only on the form of $A(y)$, as we will see
below.

\vspace{.5cm}

When  $\Lambda$ is positive, the solution for  
 $a_0(t)$ is the same as above, but  
 $A(y)$ becomes different.
One has
\beq
 A(y)={\sqrt{\lambda}\over \mu_d}\sin[\mu_d(y_H-|y|)],
          \label{desit}
\eeq 
\beq
  \mu_d=\sqrt{\Lambda/6}, \qquad \sin(\mu_d y_H)=\mu_d/\sqrt{\lambda}.
               \label{const1}
\eeq
Here $y_H$ represents the position of the horizon in the bulk $dS_5$, where
there is no spatial boundary as in $AdS_5$.
This configuration represents a brane with $dS_4$ embedded in the
bulk $dS_5$ at $y=0$. The $Z_2$ symmetry is also imposed. 

\vspace{.5cm}

Related to this solution for positive $\Lambda$,
we comment on an another form of solution \cite{NOd} which can be solved
by the following ansatz,
\beq
  ds^2= A^2(y)a_0^2(T)(-dT^2+\gamma_{ij}dx^idx^j)+dy^2, \label{nojodi}
\eeq
where we take as $a(T,y)=n(T,y)=a_0(T)A(y)$.
In this case, the solution is obtained for $k=1$ as
\beq
  a_0(T)={l\over \sqrt{\lambda}}{1\over\cos(T)},\qquad 
       A(y)={\sqrt{\lambda}\over \mu_d}\sin[\mu_d(y_H-|y|)], \label{nojodi2}
\eeq
where $A(y)$ is the same as in Eq.~(\ref{desit}) and $a_0$
has a different form. But we can see that this metric is transformed to
(\ref{desit}) by a coordinate transformation from $T$ to $t$ by
$dT/dt=a_0(t)$. So the properties of this metric will be the same
as of the metric given above. 
Similar coordinate transformations for the other solutions given above
would lead to different form of the solutions, but we will not discuss this
further here.

\subsection{Solutions for $\lambda <0$}

Another type of time-dependent solution
occurs for $\lambda <0$.
In this case, we obtain the $AdS_4$ brane 
only 
with 
$k=-1$ as seen from Eq.(\ref{Einstein2}). The solution is obtained as
\beq
 a_0(t)={1\over \sqrt{-\lambda}}\sin(\sqrt{-\lambda}t), \quad 
   A(y)= {\sqrt{-\lambda}\over\mu}\cosh(\mu[c- |y|]), \, \label{metrica42}
\eeq
where $c$ is a constant. 
(We cannot get this solution for positive $\Lambda$ since $\lambda$ is then 
 positive, as is seen from Eq.~(\ref{4cos}).)
This solution, however, would not represent
our universe since we cannot see the graviton as a massless field in this
AdS brane world \cite{KR}. 

\vspace{.5cm}
In any case, it would be important to observe the localization of various
fields,
especially the graviton,
which is needed in the brane-world in order to get a  realistic theory.
In the next section, we discuss this problem.

\section{Localization of gravity}

The case of $\lambda=0$ is well known and studied widely, 
and it is known that there is no normalizable zero-mode for $\lambda<0$.
So we discuss here the case of $\lambda>0$, where we can get solutions
for both $\Lambda>0$ and $\Lambda<0$.
Consider the perturbed metric $h_{ij}$ in the form
\beq
 ds^2= -n^2(t,y)dt^2+a^2(t,y)[\gamma_{ij}(x^i)+h_{ij}(x^i)]dx^{i}dx^{j}
           +dy^2  \,. \label{metricape}
\eeq
We are interested in the localization of the traceless transverse
component, which represents the graviton on the brane, of the perturbation.
It is projected out by the conditions, $h_i^i=0$ and
$\nabla_i h^{ij}=0$, where $\nabla_i$ denotes the covariant derivative
with respect to the three-metric $\gamma_{ij}$ which is used to raise
and lower the three-indices $i,j$. The transverse and traceless part
is denoted by $h$ hereafter for simplicity.

\vspace{.5cm}
We first consider the simple case where $\Lambda<0$ 
and $\gamma_{ij}(x^i)=\delta_{ij}$ .  Then the transverse
and traceless part $h$ is projected out by
$\partial_ih^{ij}=0$
and $h^i_i=0$, where $\delta_{ij}$ is used to raise and lower the indices
${ij}$. One arrives at the following linearized equation of $h$ in terms of
the five dimensional covariant derivative $\nabla^2_5=\nabla_M\nabla^M$:
\beq
 \nabla^2_5 h=0.  \label{scalar}
\eeq
This is equivalent to the field equation of a five dimensional free scalar.
The general form of this linearized equation for
(\ref{metricape}) is given in \cite{Lan}, so we abbreviate it here.

First, consider the case of (\ref{metrica3}), where the metric is written as
\beq
  ds^2= A(y)^2(-dt^2+a_0(t)^2\delta_{ij}dx^idx^j)+dy^2. \label{metrica6}
\eeq
In this case, Eq.~(\ref{scalar}) is written by expanding
$h$ in terms of the four-dimensional continuous mass eigenstates:
\beq
 h=\int dm \phi_m(t,x^i)\Phi(m,y) \, , \label{eigenex}
\eeq
where the mass $m$ is defined by
\beq
  \ddot{\phi}_m+3{\dot{a_0}\over a_0}\dot{\phi}_m
           +{-\partial_i^2\over a_0^2}\phi_m=-m^2\phi_m , \label{masseig}
\eeq
and $\dot{}=d{}/dt$. 
For $a_0=1$, we get the usual relation, 
$-k^2=-\eta^{\mu\nu}k_{\mu}k_{\nu}=m^2$, where
$\phi=e^{i k_{\mu}x^{\mu}}$,  $m$ representing the four-dimensional mass.
So we can consider $m$ as the mass of the field on the brane.
The explicit form of the solution of Eq.~(\ref{masseig}) is not shown here
since it is not used hereafter.
The equation for $\Phi(m,y)$ is obtained as 
\beq
  {\Phi}''+4{A'\over A}{\Phi}'
           +{m^2\over A^2}\Phi=0 , \label{warp}
\eeq
where ${}'=d{}/dy$.

Before considering the solution of Eq.~(\ref{warp}), we note that this
equation
can be written in a "supersymmetric" form as
\beq
  Q^{\dagger}Qu(z)=(-\partial_z-{3\over 2A}{\partial A\over \partial z})
(\partial_z-{3\over 2A}{\partial A\over \partial z})u(z)=m^2u(z),
\label{warp2}
\eeq
where $\Phi=A^{-3/2}u(z)$ and $\partial z/\partial y=\pm A^{-1}$. So the 
eigenvalue $m^2$ should be non-negative, i.e., no tachyon in four dimension.
Then the zero mode $m=0$ is the lowest state which would be localized
on the brane. The localization is seen by solving the one-dimensional 
Schr\"{o}dinger-like equation in the $y$-direction with the eigenvalue $m^2$,
in the form of Eq.(\ref{warp3}). The potential $V(z)$ in Eq.~(\ref{warp3}) is 
determined by $A(y)$, and it should contain a $\delta$-function attractive
force at the brane position to trap the zero-mode of the bulk graviton.
Another
condition for the localization is the existence of a normalizable state
for the wave function of $m=0$ eigenvalue.
For the solution (\ref{metrica4})
with $\gamma_{ij}=\delta_{ij}$, the discussions are also given in
\cite{LMW,KR}.

\vspace{.5cm}

For the cases of $k=\pm 1$, some of the solutions given above are 
also considered in \cite{KR}, especially for 
$\lambda <0$ and $k=-1$ ($AdS_4$ brane). An
 interesting feature is 
 that the graviton on the brane in this case might be massive
\cite{KR,Por}. We will
 not discuss this point further here.
The procedure to examine the localization for $k=\pm 1$ is parallel to the
procedure for $k=0$. 
The metric perturbation taken as in Eq.~(\ref{metricape}) is projected out
by the condition $\nabla_i h^{ij}=0$ applied to the transverse
component, and its traceless part
is denoted by $h$ as above. It satisfies  Eq.~(\ref{scalar}). By expanding
it as in Eq.~(\ref{eigenex}),
one obtains the equation for $\phi_m(t,x)$:
\beq
  \ddot{\phi}_m+2{\dot{a_0}\over a_0}\dot{\phi}_m
           +{-\nabla_i^2+2k\over a_0^2}=-m^2\phi_m . \label{masseig2}
\eeq
Similarly to the case of Eq.~(\ref{masseig}),
it is easy to see that $m$ corresponds to the  mass on the brane.
And we obtain
the same equation  (\ref{warp}) as before, for $\Phi(m,y)$. So in the 
solution of the above equation the mass has to obey the restriction
$m^2\geq 0$.

\vspace{.5cm}
For any solution, Eq.(\ref{warp}) can be rewritten in terms of $u(z)$, which
is defined by $\Phi=A^{-3/2}u(z)$ as given above, as follows:
\beq
 [-\partial_z^2+V(z)]u(z)=m^2 u(z) , \ \label{warp3}
\eeq
where the potential $V(z)$ depends on $A(y)$ as
\beq
 V(z)={9\over 4}(A')^2+{3\over 2}AA''
\eeq 
Hereafter, we examine the spectrum of $u(z)$ for various solutions given
above.

\vspace{.5cm}
First, we discuss the solution (\ref{desit}) 
obtained for $\Lambda >0$. In this case,
$$z=\rm{sgn}(y)(\lambda)^{-1/2}\ln(\cot[\mu_d(y_H-|y|)/2])$$
where $V(z)$ is expressed as
\beq
 V(z)= {15\over 4}\lambda [-{1\over\rm{cosh}^2(\sqrt{\lambda}z)}+{3\over 5}]
      -{\kappa^2\tau\over 2}\delta(|z|-z_0) , \label{pot2}
\eeq
\beq
 z_0={1\over\sqrt{\lambda}}\rm{arccosh}({\sqrt{\lambda}\over \mu_d}).
\eeq
Here $z_0$ corresponds to $y=0$, the position of the brane. (Note that 
$\sqrt{\lambda}/\mu_d \ge 1$, because of Eq.~(5).)
We notice several points with respect to this potential. 
\begin{enumerate}
\item One sees the presence of an  attractive $\delta$-function force 
at the brane for $\tau >0$.
Then one would expect one bound state on the brane, and it should
be the ground state, i.e., the zero-mode of $m=0$. 
\item The potential monotonically increases with $z$ and approaches 
\beq
      V_{\infty}={9\over 4}\lambda
\eeq
at $z=\infty$ or at the horizon $y=y_H$. (See Fig.1.) Then one might suppose
the existence of discrete Kaluza-Klein (KK) modes
in the range $0<m<V_{\infty}$. However, as will be shown below, 
there is no such  discrete mode
except from the zero-mode,  $m=0$.
\begin{figure}[htbp]
\begin{center}
\voffset=15cm
  \includegraphics[width=8cm,height=7cm]{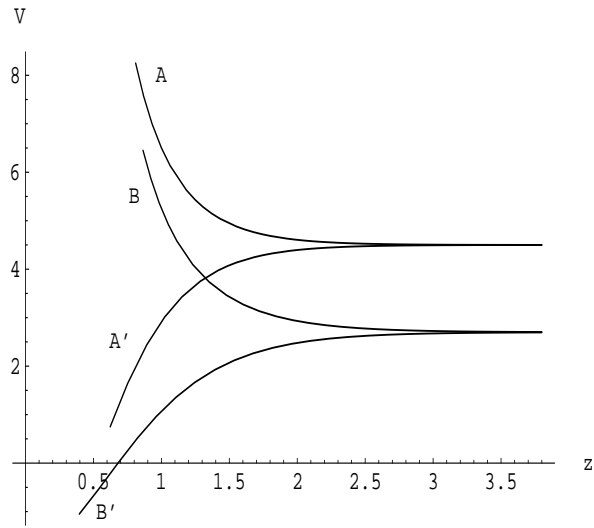}
\caption{The curves $A'(A)$ and $B'(B)$ show the finite part of $V(z)$ given
in (\ref{pot2}) ((\ref{pot3})) for the parameters $\lambda=2$ and 
$\lambda=1.2$ respectively with $\mu_d=1 (\mu=1)$. The left hand end-points
of each curve represents $V(z_0)$, where $\delta$-function attractive 
potential appears.}
\end{center}
\end{figure}
\item For $m>V_\infty$, there appear the continuum KK modes. Due to this lower
bound of the continuum spectrum of KK modes, the shift from the Newton law 
on the brane is qualitatively different from the case of the RS brane, where
the continuum KK modes are observed with $m>0$.
\item The potential takes its minimum,
\beq
 V(z_0)={5\over 8}\lambda({18\over 5}-{\Lambda\over\lambda}),
\eeq
at $z=z_0$, which is the left-hand end point of curves $A'$ 
and $B'$ in Fig.1. The value of $V(z_0)$
should be non-negative to confine the zero mode on the brane. This
requirement leads to the condition 
\beq
   \Lambda \leq ({\kappa^2\tau\over 2})^2 . \label{const2}
\eeq
This constraint is necessary for $\Lambda >0$ since $V(z_0)$ is always
positive for $\Lambda <0$,
where volcano type potential is realized although the tail
of the mountain does not approach zero but instead $V_{\infty}=9\lambda/4$
(see  curves $A$ and $B$ in Fig.1).
\item When one considers the above constraints (\ref{const2}) and
(\ref{const1}), the distance between the horizon and the brane is
restricted to be
\beq
      y_H\leq {1\over \mu_d}\sin^{-1}(\sqrt{{3\over 5}}) . \label{const3}
\eeq
\item If $\lambda$ is large, implying that also $\sqrt{\lambda}/\mu_d \gg
1$, then the
 position coordinate of the brane approaches  $z_0=\lambda^{-1/2}\ln
(2\sqrt{\lambda}/\mu_d)$,
 which is a small quantity.

\end{enumerate}

The first and second points above are explicitly examined through the
solution of
Eq.~(\ref{warp3}), which is given as
\beq
 u(z)=c_1X^{-id} {}_2F_1(a,b;c;X)+c_2X^{id}{}_2F_1(a',b';c';X), \label{solds}
\eeq
where $c_{1,2}$ are  constants of integration and
\beq
 X={1\over\rm{cosh}^2(\sqrt{\lambda}z)}, \quad 
      d={\sqrt{-9+4m^2/\lambda}\over 4},   \label{para1}
\eeq
\beq
  a=-{3\over 4}-id, \quad b={5\over 4}-id, \quad c=1-2id, \label{para2}
\eeq
\beq
  a'=-{3\over 4}+id, \quad b'={5\over 4}+id, \quad c'=1+2id. \label{para3}
\eeq
Here ${}_2F_1(a,b;c;X)$ denotes the Gauss's hypergeometric function.
It follows from this solution that $u(z)$ oscillates 
with $z$ when $m>3\sqrt{\lambda}/2$,
where the continuum KK modes appear. While for $m<3\sqrt{\lambda}/2$,
$u(z)$ should decrease rapidly for large $z$ since the mode in this region
should be a bound state. Then one must take $c_2=0$, and this solution must
satisfy the boundary condition at $z=z_0$,
\beq
 u'(z_0)=-{\kappa^2\tau\over 4}u(z_0),  \label{boundzero}
\eeq
because of the $\delta$-function in the potential.
However, we find such a state, which satisfies the above condition, 
only at $m=0$ as shown in  Fig.2.
\begin{figure}[htbp]
\begin{center}
\voffset=15cm
   \includegraphics[width=8cm,height=7cm]{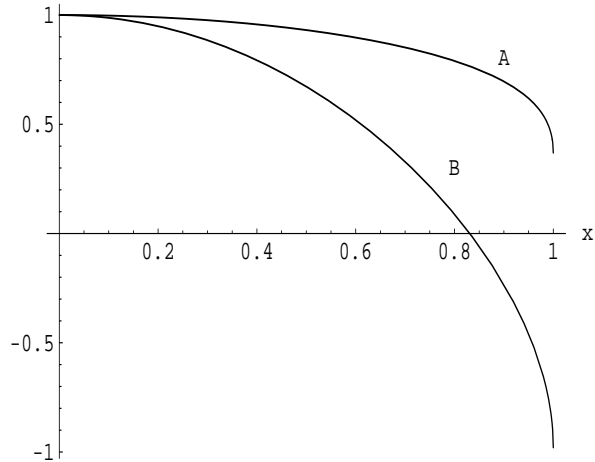}
 \caption{The curves A and B show $u(z_0)$ and $-4u'(z_0)/(\kappa^2\tau)$
respectively for $\lambda=5/3$ and $\mu_d=1$.
Mass $m$ is parameterized as $x=m/({2\over 3}\sqrt{\lambda})$. 
The two curves coincide only at $x=0$, i.e., at $m=0$.}
\end{center}
\end{figure}
Then there is no other bound state than the zero-mode, which is confined
on the brane, and the remaining mass eigen-modes, $m>V_{\infty}$, are the 
continuum KK modes with the lower mass-bound.

\vspace{.5cm}
The next point to be shown for
the localization is the normalizability of this zero-mode
in the sense
\beq
 \int_0^{y_0}dy A^2(y)\Phi^2(0,y) < \infty,   \label{norm}
\eeq
for the zero-mode solution $\Phi(0,y)$ which
is obtained as
\beq
 \Phi(0,y)=\tilde{c}_1{\cos(\mu_d[y_H-|y|]) \over 3\sin^3(\mu_d[y_H-|y|])}
           (2\sin^2(\mu_d[y_H-|y|])+1)+\tilde{c}_2,   \label{zeromo}
\eeq
where $\tilde{c}_{1}$ and $\tilde{c}_{2}$ are  integral constants. This
solution
must satisfy the boundary condition, $\Phi'(0,y=0)=0$, so $\tilde{c}_1=0$ and
$\Phi(0,y)=\tilde{c}_2$. Then the condition (\ref{norm})
is satisfied. This problem can also be studied in terms of the propagators
in the bulk space \cite{GN,Gidd}, 
but it is difficult to present such analysis 
here. 

\vspace{.3cm}
Next is the case of $\Lambda <0$ and $\lambda >0$. In this case,
$A(y)$ is given as follows,
\beq
 A(z)={\sqrt{\lambda}\over \mu\rm{sinh}(\sqrt{\lambda}|z|)},
\eeq
where
$z=\rm{sgn}(y)(\lambda)^{-1/2}\ln(\rm{coth}[\mu(y_H-|y|)/2])$ and
$V(z)$ is expressed as
\beq
 V(z)= {15\over 4}\lambda [{1\over\rm{sinh}^2(\sqrt{\lambda}z)}+{3\over 5}]
      -{\kappa^2\tau\over 2}\delta(|z|-z_0) . \label{pot3}
\eeq
The position of the brane $z_0$ ($y=0$) is given by
\beq
 z_0={1\over\sqrt{\lambda}}\rm{arcsinh}({\sqrt{\lambda}\over \mu}).
\eeq
In contrast to the case of
$\Lambda>0$, the potential (\ref{pot3}) has a volcano form, although this
mountain ends at $z=\infty$ with $V_{\infty}$ as mentioned above.
(See Fig.1., curves $A$ and $B$.)
However, the discussions are parallel and the conclusions for the
localization are the same except for the constraint on $\Lambda$. 
Namely, (i) Only the zero mode ($m=0$)
is bounded and there is no other bound state in the expected region,
$0<m<3\sqrt{\lambda}/2$. (ii) Only the continuum KK mode
for $m>3\sqrt{\lambda}/2$ occurs.
Also, it is to be noted from Eq.~(58) that in the limiting case when
$\lambda \rightarrow 0$, $z_0$ approaches $z_0=1/\mu=\sqrt{-6/\Lambda}$,
which is independent of $\tau$.

\vspace{.3cm}

As in the previous case, these results can be assured by the explicit
solution for $u(z)$
with the above potential (\ref{pot3}). It is obtained as
\beq
 u(z)=\bar{c}_1Y^{-id} {}_2F_1(a,b;c;-Y)+\bar{c}_2Y^{id}{}_2F_1(a',b';c';-Y), 
     \label{solds2}
\eeq
where $\bar{c}_{1,2}$ are integration constants and
\beq
 Y={1\over\rm{sinh}^2(\sqrt{\lambda}z)}.
\eeq
Here $(a,b,c), (a',b',c')$ and $d$ are the same as those given in Eqs.~
(\ref{para1}) $\sim$ (\ref{para3}).
The function $u(z)$ oscillates 
with $z$ for $m>3\sqrt{\lambda}/2$,
where the continuum KK modes appear. For  $m<3\sqrt{\lambda}/2$,
$u(z)$ decreases rapidly for large $z$ as before. 
Then one must take $\bar{c}_2=0$, and this solution must
satisfy the boundary condition (\ref{boundzero}) at $z=z_0$.
Then we find the above condition only at $m=0$.
This is shown in  Fig.3.
\begin{figure}[htbp]
\begin{center}
\voffset=15cm
   \includegraphics[width=8cm,height=7cm]{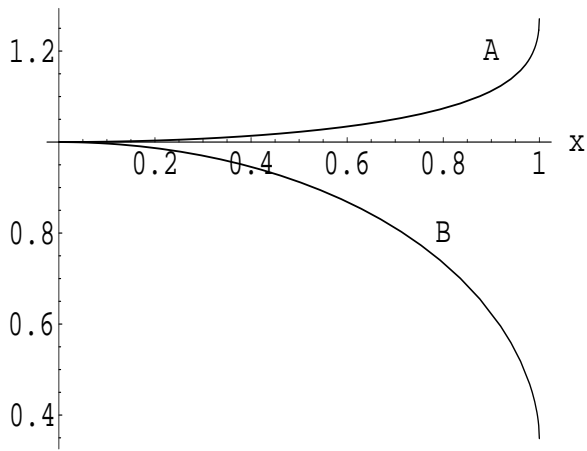}
 \caption{The curves A and B show $u(z_0)$ and $-4u'(z_0)/(\kappa^2\tau)$
respectively for $\lambda=10/3$ and $\mu=1$.
And mass $m$ is parameterized as $x=m/({2\over 3}\sqrt{\lambda})$. 
The two curves coincide only at $x=0$, i.e., at $m=0$.}
\end{center}
\end{figure}

 
As a result, the localization of zero-mode occurs both for positive and
negative $\Lambda$, but the value of $\Lambda$ is restricted as
\beq
   -{1\over 6}\leq{\Lambda \over \kappa^4\tau^2} <0,
             \label{constraint1}
\eeq
for $AdS_5$ bulk space, and
\beq
   0< {\Lambda \over \kappa^4\tau^2}\leq {1\over 4},
             \label{constraint2}
\eeq
for $dS_5$ bulk.
The constraint (\ref{constraint1}) comes from the positivity of
$\lambda$,
and (\ref{constraint2}) represents Eq.~(\ref{const2}), which is required
in the
case of positive $\Lambda$. So we conclude that the brane with
a tiny positive cosmological constant should be embedded in the 5d bulk
space with $\Lambda$ in the range given above in order to localize
gravity.

\vspace{.5cm}



\section{Cosmological implications to the $C/a^4$ term }

The product anzatz (\ref{metrica3}) is essential in our
present analyses, and this anzatz is equivalent to setting as
$C=0$ in the solutions 
(\ref{fullsol1}) and (\ref{fullsol}). 
In this section, we show that this setting $C=0$ is reliable and 
applicable to the early inflationary 
epoch and the present mini-inflationary one. We also present some
cosmological implications to the coefficient $C$.
The term $C/a_0^4$ in Eq. (\ref{Einsteina0}) is called as the dark radiation.
From the viewpoint of the AdS/CFT correspondence \cite{Gub,Pad},
the dark radiation can be regarded as
CFT radiation. We then show the relation of $C$ to temperature of CFT 
radiation. 
Also, the sub-class of viscous cosmology is briefly dealt with.
It is known from non-viscous theory that radiation is proportional to $a^{-4}$.
We point out how the presence of a bulk viscosity
destroys the simple property. 



\subsection{Justification of $C=0$ and temperature of CFT}

 The evolution of our universe consists of four epochs;
 (1) the early inflationary epoch, (2) the radiation dominated epoch, 
 (3) the matter dominated epoch and (4) the present mini-inflationary epoch.
Epoch (4) is supported by recent observations of Type Ia supernovae and 
the cosmic microwave background \cite{garnavich,perlmutter,wang}, since they  
indicate that our universe is accelerating.
The energy density of our universe is dominated by the effective 
cosmological constant in epochs (1) and (4).
The present analyses are then applicable for these epochs, 
as long as the dark radiation $C/a_0(t)^4$ is negligible in the solutions
(\ref{fullsol1}) and (\ref{fullsol}).
The energy density $\rho_{DR} \equiv C/a_0(t)^4$ of the dark radiation 
is constrained by Big-Bang Nucleosynthesis (BBN) \cite{Yahiro}. The result
is $\rho_{DR}/\rho_{r} \lap 0.05 $ at BBN epoch, 
where $\rho_{r}$ is the radiation energy density. 
The ratio is almost time indepentent, mentioned below.
The ratio then persists in epochs (1) and (4) 
where even $\rho_{r}$ is negligible compared with $\lambda$.
The dark radiation is thus 
negligible in epochs (1) and (4). The product ansatz (\ref{metrica3}) is 
then true there, since it can be obtained by taking $C=0$ in (\ref{fullsol1}) 
and (\ref{fullsol}).


The dark radiation can be related to the CFT radiation 
from the viewpoint of the AdS/CFT correspondence \cite{Gub,Pad}.
The radiation energy density at the BBN era is obtained as 
$\rho_r=g(T_{BBN}) T_{BBN}^4 \pi^2/30$, where $g(T_{BBN})$ is 
the effective number
of relativistic degrees of freedom at temperature $T_{BBN}$ 
in the BBN era. 
In the standard model, $g(T_{BBN})=10.75$. 
In the four-dimensional conformal symmetric 
Yang-Mills theory with ${\cal N}=4$, the corresponding effective number 
$g_{CFT}$ is $15(N^2-1)*3/4$ for SU(N) gauge, where the factor $3/4$
is an effect of strong interactions \cite{Gub}.
The CFT radiation (dark radiation) 
energy density is 
then $\rho_{DR}=\rho_{CFT}=g_{CFT} T_{CFT}^4 \pi^2/30$. The temperature $T_{CFT}$ of
CFT 
differs from real temperature $T$, since CFT has no coupling with 
ordinary fields except graviton.
Hence, we obtain
\beq
     \delta \equiv \frac{g_{CFT} T_{CFT}^4}{g(T_{BBN})T_{BBN}^4} \lap 0.05, 
            \label{delta}
\eeq
and then
\beq
     \frac{T_{CFT}}{T_{BBN}} \lap \frac{1}{\sqrt{5N}} 
\eeq
for large $N$.
This ratio is estimated in the BBN era, but it is almost time independent 
since real and CFT temperatures, $T$ and $T_{CFT}$, are 
almost proportional to $1/a_0(t)$.
>From a theoretical point of view, 
$N$ should be large. For such large $N$, $T_{CFT}$ is proportional to 
$1/\sqrt{N}$ and then much smaller than $T$.

It is likely that both $\rho_{CFT}$ and $\rho_{r}$ are generated 
in the reheating era just after the early inflation. 
The constraint (\ref{delta}) for $\delta$ is somewhat modified in the era,
since the effective number $g(T_{*})$ at the era is $106.75$ 
in the standard model 
and different from $g(T_{BBN})$. 
A precise estimate is possible on the basis of entropy conservation,
$a_0(t_{BBN})^3 g(T_{BBN}) T_{BBN}^3 = a_0(t_{*})^3 g(T_{*}) T_{*}^3$ and
$a_0(t_{BBN})^3 g_{CFT} T_{CFT}^3 = a_0(t_{*})^3 g_{CFT} T_{CFT*}^3$, 
where $*$ stands for the reheating era.  For example, $T_{CFT*}$ is the
CFT temperature in the era. The ratio $\delta(t_*)$ 
at the reheating era is 
\beq
     \delta(t_*) \equiv \frac{g_{CFT} T_{CFT*}^4}{g(T_{*})T_{*}^4} 
          \lap 0.05 ( \frac{g(T_{*})}{g(T_{BBN})} )^{1/3}
          \sim 0.11 . 
            \label{delta*}
\eeq
This ratio can be regarded as a ratio of the coupling of inflaton with CFT 
fields to that with ordinary fields. The former coupling is thus 
at least an order of magnitude smaller than the latter. 

\subsection{On viscous cosmology}

In cosmological theory, the cosmic fluid with four-velocity $U^\mu$ is most
often taken to be ideal, {\it i.e.}, to be nonviscous. From a
hydrodynamical viewpoint this idealization is almost surprising, all the
time that the  viscosity property is so often found to be of great physical
significance in ordinary hydrodynamics. As one might expect, the  viscosity
concept has gradually come into more use in cosmology in recent years, and
it may seem to be pertinent to deal briefly with this topic here also,
emphasizing in particular the close relationship between viscous theories
and  nonconformally invariant field theories.

Consider the fluid's energy-momentum tensor (cf., for instance,
Ref.~\cite{brevik02}):
\begin{equation}
T_{\mu \nu}=\rho U_\mu U_\nu+(p-\zeta \theta)h_{\mu \nu}-2\eta \sigma_{\mu
\nu},
\end{equation}
where $\zeta$ is the bulk viscosity and $\eta$ the shear viscosity,
$h_{\mu\nu}=(g_{\mu\nu}+U_\mu U_\nu)$ the projection tensor, 
$\theta={U^\mu}_{;\mu}$ the scalar expansion, and $\sigma_{\mu\nu}$ the
shear tensor. Here we ought to notice that under normal circumstances in
the early universe the value of $\eta$ is enormously larger than the value
of $\zeta$. As an example, if we consider the instant $t=1000$ s  after Big
Bang, meaning that the universe is in the plasma era and is characterized
by ionized H and He in equilibrium with radiation, one can estimate on the
basis of kinetic theory \cite{brevik94} that the value of $\eta$ is about
$2.8 \times 10^{14}\,{\rm g\,cm^{-1}\,s^{-1}}$, whereas the value of
$\zeta$ is only about $7.0 \times 10^{-3}\,{\rm g \,cm^{-1}\, s^{-1}}$.
Even a minute deviation from isotropy, such as we encounter in connection
with a Kasner metric, for instance, would thus be sufficient to let the
strong shear viscosity come into play. However, let us leave this point
aside here, and follow common usage in taking the universe to be perfectly
homogeneous and isotropic. Then, $\eta$ can be ignored and we obtain, for a radiation dominated universe with
$p=\rho/3$ \cite{brevik02},
\begin{equation}
\frac{d}{dt}(\rho_r\,a^4)=\zeta\, \theta^2\,a^4,
\end{equation}
where $a$ is the conventional scale factor. The important point in our
context is the following: In the presence of viscosity the content $\rho_r$
of radiation energy is no longer proportional to $a^{-4}$. We recall that
energy density terms of the form $C/a^4$ were found above, both in the
five-dimensional theory (Eq.~(11)), and on the brane (Eq.~(15)).
Physically, the inclusion of even a simple bulk viscosity coefficient means
a violation of conformal invariance (like mass terms).

\vspace{.5cm}

\section{Concluding remarks}

We have examined the localization of gravity on various cosmological
branes with a non-trivial curvature. The cosmological constant $\Lambda$ 
in the bulk space 
is considered for both negative and positive values. The latter
case ($\Lambda >0$) has recently attracted interest in connection with the
proposed dS/CFT 
correspondence. In both cases, we find a localized zero mode of the
gravity fluctuation for a restricted region of $\Lambda$. 

For negative $\Lambda$, the bulk is asymptotically $AdS_5$ and the solution
known as RS brane with flat four-dimensional metric can be obtained by 
fine-tuning 
the parameters. On this brane, the localization of  gravity has been
demonstrated by number of previous works. 
Different choices for the parameters in our model can lead to positive
as well as negative values for
the four-dimensional cosmological
constant ($\lambda$). In the case of negative $\lambda$, $AdS_4$ space is
obtained as 
a solution for the brane-world with $k=-1$, and we cannot obtain normalizable
zero-mode of gravity fluctuations. This
implies that we cannot observe the usual four-dimensional
gravity on this brane. There is a normalizable zero-mode
in the case of positive $\lambda$, and we find that this mode is confined
on the brane. In contrast to the case of $\lambda=0$, 
 the continuous mass of
the KK modes has a lower bound which is proportional to $\lambda$. We also
demonstrated that there is no bound-state on the brane below this lower bound 
other than the zero-mode, the four-dimensional graviton.

For positive $\Lambda$, the bulk is asymptotically $dS_5$ and the brane
is realized only for positive $\lambda$. Although the graviton could be 
localized 
on the brane, the situation in the bulk
is different from the case of negative $\Lambda$. 
In fact, we find a critical value
of $\Lambda$, below which we can see the localized graviton, which could lead
to the usual Newton law, on the brane. Above
this critical value, the zero mode expands in the fifth dimension
and one cannot see the four-dimensional graviton anymore on the brane. 
The KK mode has the same lower mass-bound as in the case of  negative 
$\Lambda$,
but the wave function is different due to the part dependent on the fifth 
coordinate. So, the contribution to the shift from the Newton law will be 
discriminated from the case of  negative $\Lambda$.

Also from the smallness of the presnt $\lambda$, 
the value of the positive $\Lambda$ should be bounded, 
but we cannot reject the possibility of
positive $\Lambda$ in considering our brane-world.  Our current
conclusion  is that the value of $\Lambda$ can be restricted
to the region given in Eqs.~(\ref{constraint1}) and (\ref{constraint2}).
Detailed analysis is necessary in order to restrict the parameters region
in a realistic brane-world. Moreover, such an analysis may indicate 
if the bulk space should be de Sitter or anti-de Sitter one.

  All the analyses mentioned above are based on the assumption 
that the radiation part
of CFT on the boundary is negligible. 
This assumption is reliable for the early inflationary 
epoch and the present mini-inflationary one. The smallness of the part 
indicates that so is the temperature of CFT. 
The possible applicability of our results to viscous cosmology is also given.

\end{document}